\documentclass{actasen}

\usepackage{diagbox,multirow}

\usepackage{cite}
\newcommand{\upcite}[1]{\textsuperscript{\textsuperscript{\cite{#1}}}}

\begin{document}

%%ÉèÖÃÊ×Ò³Ò³Âë
\setcounter{page}{1}

\Volume{*}{*}% Äê¡¢¾í

%%ҳüÉèÖÃ

\runheading{An Tao et al.}%%%%%%%%%%×÷ÕßÃû×Ö

\title{Radio Frequency Interference Mitigation$^{\dag}~ \!^{\star}$}%%%%%%%%%%%%ÎÄÕ±êÌâ

\footnotetext{$^{\dag}$ 
National Key Research and Development Program of China (2016YFE0100300), Strategic Priority Research Program of the Chinese Academy of Sciences (XDB23000000), NSFC Research Fund for International Young Scientists (11650110438).

Received *--*--*; revised version *--*--*

$^{\star}$ A translation of {\it Acta Astron. Sin.~}
Vol. *, No. *, pp. *, * \\
\hspace*{5mm}$^{\bigtriangleup}$ antao@shao.ac.cn\\

\noindent 0275-1062/01/\$-see front matter $\copyright$ 2011 Elsevier
Science B. V. All rights reserved. %%

\noindent PII: }

\enauthor{An Tao$^{\bigtriangleup}$$^{1,2}$\hs\hs Chen
Xiao$^{1}$\hs\hs 
Mohan Prashanth$^{1}$\hs\hs Lao Bao-Qiang$^{1}$}
{\up{1} Shanghai Astronomical Observatory, the Chinese Academy of Sciences, Nandan Road 80, Shanghai 200030 
\up{2} 2 Key Laboratory of Radio Astronomy, Chinese Academy of Sciences, 210008 Nanjing, China \\
}%%%%%%%%%%%%%%%%%%%%%%%% Á¥Êôµ¥Î»

\abstract{Radio astronomy observational facilities 
are under constant upgradation and development to achieve better capabilities including increasing the time and frequency resolutions of the recorded data, and increasing the receiving and recording bandwidth.  
As only a limited spectrum resource has been allocated to radio astronomy by the International Telecommunication Union,  
this results in the radio observational instrumentation 
being inevitably exposed to undesirable radio frequency interference (RFI) signals which originate mainly from terrestrial human activity and are becoming stronger with time. RFIs degrade the quality of astronomical data and even lead to data loss. The impact of RFIs on scientific outcome is becoming progressively difficult to manage. 
In this article, we motivate the requirement for RFI mitigation, and review the RFI characteristics, mitigation techniques and strategies. Mitigation strategies adopted at some representative observatories, telescopes and arrays are also introduced. We also discuss and present advantages and shortcomings of the four classes of RFI mitigation strategies, applicable at the connected causal stages: preventive, pre-detection, pre-correlation and post-correlation. The proper identification and flagging of RFI is key to the reduction of data loss and improvement in data quality, and is also the ultimate goal of developing RFI mitigation techniques. This can be achieved through a strategy involving a combination of the discussed techniques in stages. Recent advances in high speed digital signal processing and high performance computing allow for performing RFI excision of large data volumes generated from large telescopes or arrays in both real time and offline modes, aiding the proposed strategy. }


\maketitle

\section{Introduction} %%%%%%%%%%%%%%%Ò»¼¶±êÌâ

Radio frequency interference (RFI) with relation to radio astronomy can be broadly defined as any undesirable radio signal mainly contributed to by human activity including television (TV), frequency modulation (FM) radio transmissions, communication signals including the Global Positioning System (GPS), mobile, and from airplanes, amongst others, hindering the reception of underlying faint astronomical signals\upcite{43}. Interference typically leads to both data loss and to a loss of data quality. Commercial FM broadcasts are generally in the 87.5 -- 108 MHz frequency range and are mostly narrow band persistent contaminants for astronomical signals. The TV broadcasts can be in both narrow and broad bands, spanning a wide range of frequencies and are classified as the Very High Frequencies (VHF) band, typically at 40 -- 80 MHz and 160 - 230 MHz; and the Ultra High Frequencies (UHF) band, typically at 470 -- 960 MHz. The VHF frequencies including the FM radio bands are also employed for the transmission and reception of walkie-talkie based signals used for popular civilian purposes and military communication. Typical mobile communication following the Global System for Mobile Communications (GSM) standards are operative in narrow bands around (850, 900, 1800, 1900) MHz. A comprehensive list of frequency allotments and operational regulations including allowed broadcast power from stations is presented in the framework recommended by the radio communication sector of the International Telecommunication Union (ITU)\footnote{https://www.itu.int}. The operational frequencies of the above major sources of RFI are summarized in Table \ref{Table:RFISource} and a typical spectrum of RFI occupancy at the Murchinson Widefield Array (MWA\footnote{http://www.mwatelescope.org/}) site is presented in Fig. 1\upcite{89}.
%%%%%%%%%%%%%%%%%%%%%%%
%%%%%%%%%%%%%%%%%%%%%±í¸ñ1
\begin{table}[h]
\centering
\caption{Some sources of RFI contamination at low radio frequencies}

\fns \tabcolsep 2.2mm
\begin{tabular}{ccc}
\hline
Part of RFI source &Frequency(MHz) & Characteristics/comments  \\
\hline
  Commercial FM   & (87.5 - 108)  &Narrow band; persistent      \\
      TV   & (40 - 80), (160 - 230),
 (470 ¨C 960)
  &Narrow and broad bands; \\
  & & persistent and transient       \\
      Mobile GSM   & (850, 900, 1800, 1900)  &Narrow band       \\
\hline
  Navigation satellites:GPS & (1227.60, 1575.42) & Characteristics; persistent \\
  GLONASS & 1602 & Characteristics; persistent \\
\hline
  communication satellites:
Iridium & (1618.85 - 1626.5) & Narrow and broad bands; \\
& & persistent \\
ORBCOMM & (137 - 138) & Narrow and broad bands;  \\
& & persistent \\
\hline
 \end{tabular}

\label{Table:RFISource}
\end{table}

%ͼ1
\begin{figure}[tbph]
\label{pic1}
\centering
{\includegraphics[angle=90,width=9cm]{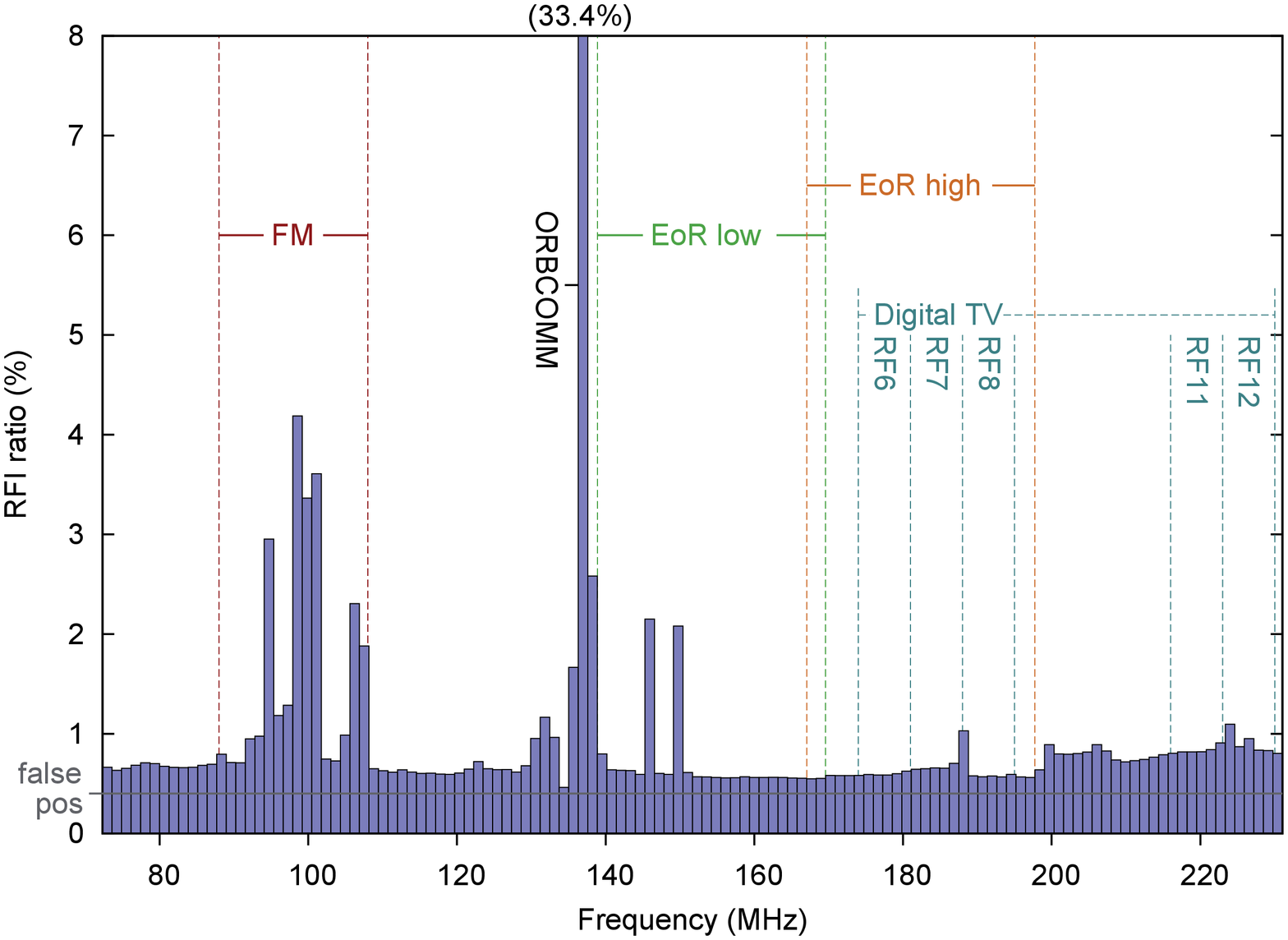}}
%\vspace{-2mm}
\caption{RFI occupancy in the low frequencies between 70 and 230 MHz (including those used for astronomical observations with the Murchinson Widefield Array; MWA in western Australia) showing contamination due to FM, a communication satellite and digital TV transmission signals. Image source: Offringa et al.\upcite{89}}
    \end{figure}

Such known sources of RFI are contaminants to narrow band frequency channels and can persist in time; in addition, RFI caused by lightning, power transmission cables, electrical fences, and other similar features which could cause transient surges are local in time but can contaminate multiple channels and hence display broadband characteristics.

Radio astronomy service (RAS) is a passive service, easily affected by the above described active services which generate interference to RAS as the ITU only allocates limited spectrum resources for radio astronomy (refer to the ITU report Techniques for mitigation of radio frequency interference in radio astronomy\footnote{ITU-R RA.2126-1, Geneva, 2007: available from http://www.itu.int/publ/R-REP/en}). Though, with an increasing bandwidth available for RAS systems, contamination from these active services is unavoidable. With the improving sensitivity of radio astronomical instrumentation for the reception of weak astronomical signals, they are often exposed to strong interference. The broadband, spread-spectrum applications for broadcasting and communication as well as of unlicensed, mass produced devices that replace the peaked, high-power signals of yore with broader signals of lower power generate signals that are not easy to remove from RAS data, thus aggravating the existing situation. Another situation encountered is the dynamic spectrum access (DSA) which allows systems to operate in spectrum slots that are unused for a period of time (e.g. boosting of widely used cognitive radio devices). The intensification of spectrum use of DSA is expected to increasingly change the character of the RFI environment, and will require the radio astronomy community to adjust its approach to reducing the impact of RFI on its data. In addition, the time variability of RFI may particularly impact time-critical astronomical observations (e.g., pulsars, fast radio bursts, radio transients). Variable, non-repetitive RFI occurring during studies of transients and pulsars may destroy critical observations that are unique and cannot be repeated. These can affect the time sequencing of pulses, and the calibration of the data. 
An example are the short duration impulses called ``Perytons" observed by the Australian Parkes telescope which was later discovered to originate from terrestrial sources and human activity (e.g. Kulkarni et al.\upcite{64}; Petroff et al.\upcite{96}).

Interference mitigation in radio astronomy is thus aimed at reducing or removing the impact of active services in bands outside those allocated to the RAS. The proper identification and flagging of RFI will help in their excision, allowing for the operation of observational facilities in environments with varying levels of RFI contamination. RFI mitigation techniques are then essential for operating in non-allocated and non-protected bands. In this article, we discuss the RFI characteristics including types of signals and their morphological properties as inferred from the time-frequency distribution of the received or calibrated signals in Chap. 2. Then, mitigation strategies adopted at the level of the observatory involving preventive and RFI reducing measures, and those that are applied on the received signal including pre-detection, pre-correlation and post-correlation are discussed in Chap. 3. We then present procedures adopted by some observatories for RFI mitigation in Chap. 4, following which we summarize the compilation in Chap. 5.

\section{RFI characteristics}

The expected RFI characteristics are used for identification, subsequent excision and possible data reconstruction. Both strong and weak RFI signatures can affect multiple related channels (broad band), thus being more distributed or only certain channels (narrow band) where it is more confined and sharper\upcite{123}. Depending on the originating source and observation time during the course of the day, it can be either transient or persistent, and measures must be taken to distinguish this from real astronomical transient sources such as pulsars, fast radio bursts (FRBs) or rotating radio transients (RRATs). In addition, RFI can occur in impulse type bursts (high amplitude, small time step) persistently, variable in time or in a transient manner. The RFI amplitudes are typically much higher compared to the underlying astronomical signal and random noise, but some weak RFI signals are almost identical to the intensity of real celestial signals such as pulsars and can easily lead to misidentification. An important distinguishing feature for the separation of terrestrial sources of RFI from the astronomical signal relates to the amount of dispersion in the ``time-frequency'' domain of the received data. Astronomical data (e.g. pulsar signals), as a result of traversing the vast interstellar or intergalactic medium are expected to be dispersed by the intervening cold plasma and gas clouds such that the amount of dispersion of frequency is based on the arrival time of the signal as $\tau  \propto {v^{ - 2}}$\upcite{105}. In comparison, terrestrial RFI signals are not or only mildly dispersed owing mainly to reflection by the ionosphere and interaction with the intervening atmosphere. 
In addition, most direct and indirect or reflected RFI transmission is mainly detected from the side-lobe coupling to active services while there is a destructive main-beam coupling from satellites and aeronautical services. Care must be taken in their classification as we might be led astray such as mentioned in Chap. 1, in the study of the nature of Perytons.

RFI properties in the integrated post-correlated data can be studied in terms of their morphologies, consisting of broadband (distributed or strongly present) and narrowband roughly horizontal and vertical envelopes, mainly caused by the earlier described terrestrial sources. In addition, the envelope can also be curved in the time-frequency domain, often caused by the Doppler shift in communication signals broadcast by overhead satellites such as the GPS (1.575 GHz, 1.228 GHz; e.g. Offringa et al.\upcite{84}). These morphological features may be either transient (closely clustered) or persistent (hence widely scattered and loosely clustered). Some examples of these RFI characteristics are presented in Figs. 2 and 3.
%ͼ2
\begin{figure}[tbph]
\label{pic2}
\centering
{\includegraphics[angle=0,width=9cm]{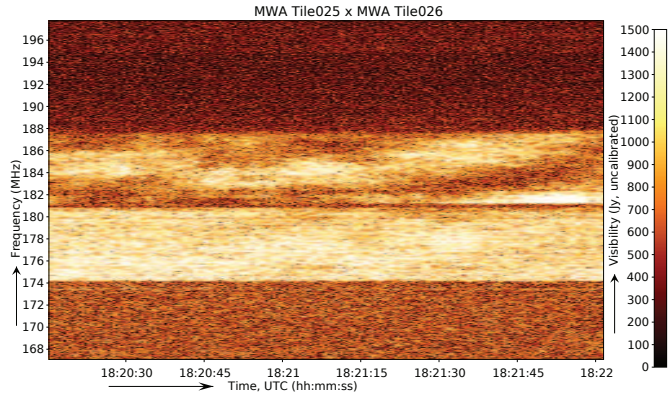}}
%\vspace{-2mm}
\caption{Persistent broadband RFI from digital TV signals in the MWA data. Image source: Offringa et al.\upcite{89}.}
    \end{figure}

%ͼ3
\begin{figure}[tbph]
\label{pic3}
\centering
{\includegraphics[angle=0,width=9cm]{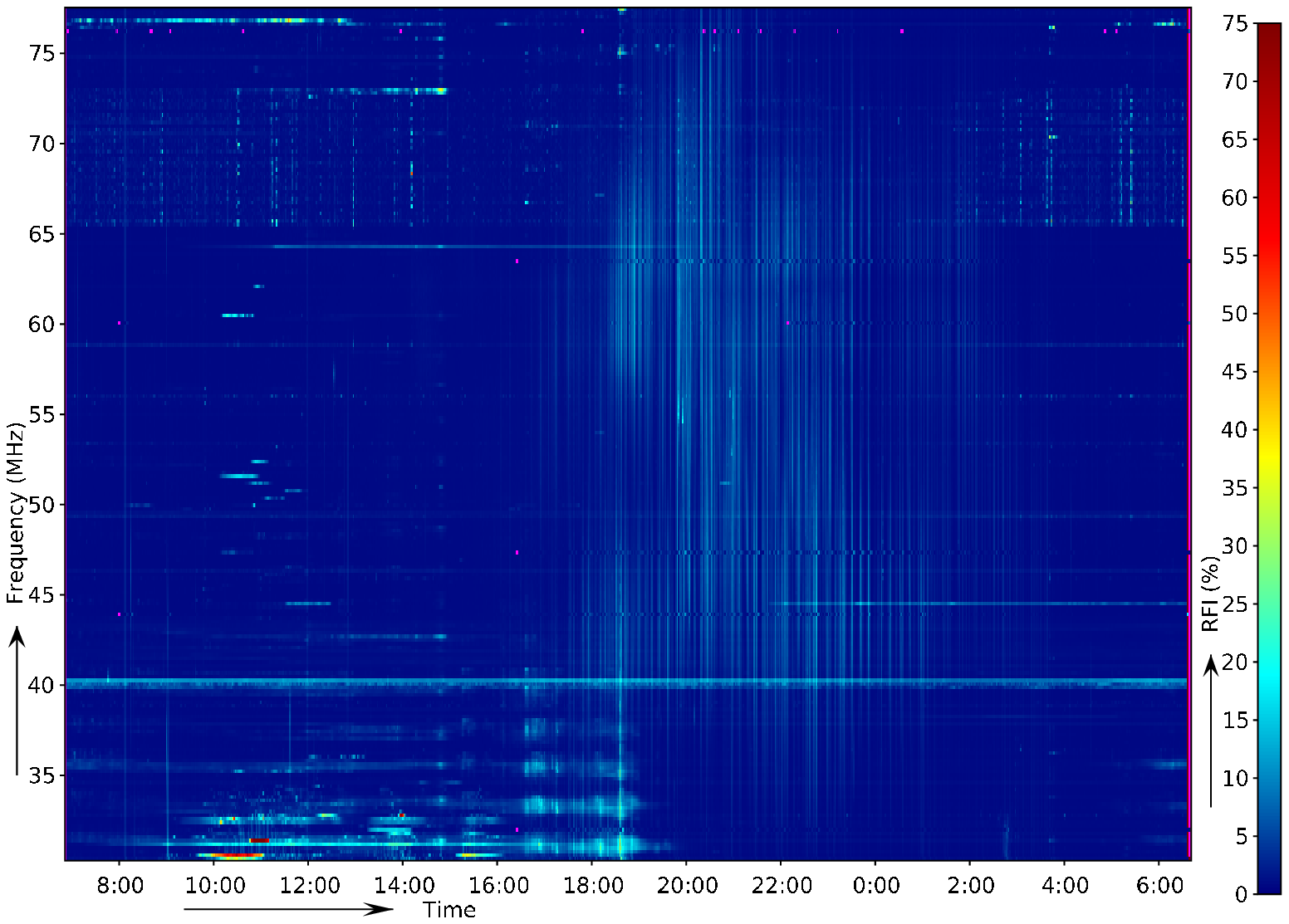}}
%\vspace{-2mm}
\caption{Narrowband and broadband RFI showing persistent and transient characteristics in the (10 - 80) MHz low frequency measurements made by the LOFAR of the Netherlands. Image source: Offringa et al.\upcite{88}.}
    \end{figure}

\section{RFI mitigation strategies}

The particular mitigation strategy adopted can be tailored to suit a specific observatory depending on its location and exposure to RFI there. Major strategies for RFI mitigation in stages can be broadly classified into (1) observatory prevention, regulatory measures adopted by the observatory during its setup and operation; (2) pre- detection, applied to the data receiver system, possibly in connection with the data-taking backend; (3) pre-correlation, including hardware or software based operations applied in real time to the incoming data before it is sent to the correlator; and (4) post-correlation, applied to integrated data or data buffers enabling offline processing. Strong RFI can be easily identified in the above four stages, while weak RFI has to be identified later in the processing chain (stage 4) after integrating the data to increase the signal-to-noise ratio.

Important considerations in the implementation of one or multiple techniques above include: (1) how to quantify the strength of RFI to set suitable thresholds? methods depend on whether the data relates to continuum or spectral line based observations; (2) how to avoid receiving false positives to ensure the accuracy of detection? (3) how to quantitatively evaluate the implementation of the RFI reduction strategies?; (4)the technical feasibility of implementation, which include hardware availability and cost, assembly and implementation time, high-performance computing costs for large-scale data, and the efforts for the maintenance of this system of techniques. A comparative review of the diverse strategies to gauge their detection efficacy and relative advantages are discussed in Fridman \& Baan\upcite{43} with applicability to both single dish telescopes and interferometric arrays. Other comprehensive overviews of these strategies have been discussed in the ITU report ITU-R RA.2126-1, Baan\upcite{10} and references therein.

The main purpose of these strategies is to minimize data loss due to RFI, by carrying out flagging and excision of RFI with minimum loss of astronomical information. Thus, the earlier an interference problem is dealt with in the processing chain the better as it results in less damage to the data, lower downstream costs, and less system complexity.

\subsection{Preventive strategies}

Preventive or pro-active strategies include the search for existing RFI free environments to set up the facility, often viable in regions sparsely populated by human settlements. The creation of a RFI-free perimeter/RFI-free zone around upcoming and existing facilities is then carried out. There is thus a requirement to advance support for policy making and implementation of regulations regarding the use of specific bands and spectrum usage in the vicinity of the observatory at a governmental level (e.g. as recommended by the ITU), which includes the prevention of the setup of transmitters in the radio quiet zone vicinity. An example is the setup and structure of such regulatory arrangements, summarized by the Committee on Radio Astronomical Frequencies (CRAF) \footnote{http://www.craf.eu/radio-quiet-zones-around-observatories/}, operational in Europe and presented in the report compiled by Cohen et al.\upcite{30}. The Figure 4 demonstrates the setup of the radio quiet zone surrounding the Five-hundred-meter Aperture Spherical radio Telescope (FAST\footnote{http://fast.bao.ac.cn/en}), the largest single-dish radio telescope located in Guizhou province, China. Any radio transmission stations or other microwave-generating equipments are prohibited in the core region of the inner 5 km region (the inner circle in Figure 4); installation and employment of any radio station with an effective power of more than 100 Watts working at 68-3000 MHz frequency range is not allowed within the central region of the 5-10 km ring. When setting up radio stations and constructing electromagnetic radiation generating facilities, the electromagnetic compatibility should be assessed and organized by provincial radio management institutions. Any devices that interfere with the normal operation of the radio telescopes may not be constructed.
%ͼ4
\begin{figure}[tbph]
\label{pic4}
\centering
{\includegraphics[angle=0,width=9cm]{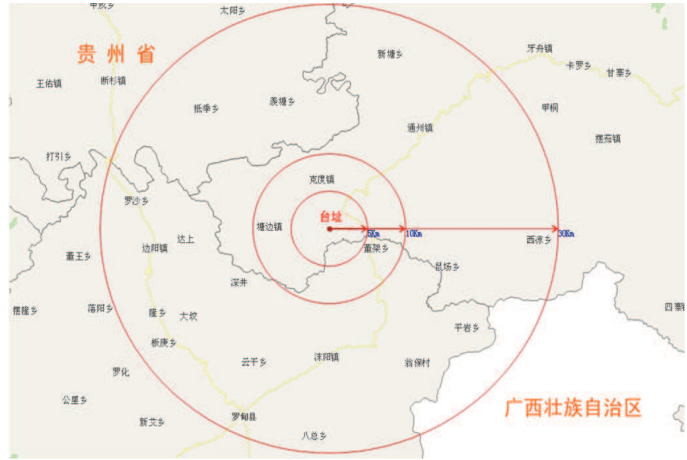}}
%\vspace{-2mm}
\caption{Illustration of the radio quiet zone of the Five-hundred-meter Aperture Spherical radio Telescope (FAST). Image courtesy of FAST Project Office.}
    \end{figure}

In addition to site selection to avoid nearby ground based radio frequency interference devices, preventive strategies at the level of the observing facility also minimize the reception of RFI contaminated data. It must be ensured that computer systems and other wireless connected devices (e.g. Wireless Local Area Networks; WLAN at 2.4 GHz) in the vicinity of the observational setup must be shielded effectively to prevent their contamination of the received signals. In addition, there is RFI originating from the electronics involved in the power supplying equipment (transmission cables, generation/transformer stations), and reception and processing instrumentation which could inadvertently produce electromagnetic emission of an undesirable nature which interferes with the received signals\upcite{126,127,128}, and has to be suitably excised especially for observations where the expected signal is weak (e.g. epoch of reionization studies). In the broad sense, the study of identification, signal attenuation and elimination methods is called electromagnetic compatibility (EMC). Under the framework of the CRAF is the case study at the Westerbork Radio Telescope (WSRT; CRAF-04-1: \upcite{74}) \footnote{http://www.craf.eu/publications/document/} where RFI contamination is identified as due to the equipment of the observatory's own devices, such as network devices (switches), computers (CPUS, monitors), read and write operations of storage disks, and receivers, and some strong interference signals from electronic monitoring equipment. 

Preventive work requires long time monitoring to obtain frequency, appearance time, azimuth and other information of known strong RFI sources, to identify and excise them, and to provide guidance on choosing the right observation windows in a day. Observations that are sensitive to RFI, such as pulsars and spectral lines, can be carried out by the operation of facility during times of the day/night when RFI contamination is minimal. The interfering signals from the satellites are unavoidable, but their frequencies and operating orbits are constant. The GPS signals are employed to accurately time-stamp all received data in order to be deleted directly in subsequent operations. In addition, there must also be a monitoring of the environment surrounding the facility to account for ionospheric changes in the received data and to track instrumental effects such as the biased increase in receiver gain upon the reception of strong signals from terrestrial sources.

\subsection{Pre-detection strategies}

Pre-detection strategies involve the installation of a bandpass or high/low pass filter in the receiver to enable frequency domain excision of strong (known) RFI  
outside the operational bands. The principle and operation of this strategy is simple. Though, shortcomings include an insertion loss with a substantial raise in the system temperature at frequencies close to a band-edge on installation of a bandpass filter. The filtering may also result in data loss for continuum observations, while it is often essential to enable spectral line observations where RFI occurs at a critical frequency within a receiver's passband. The blanking or stopping of data-acquisition process for small durations of time may also be used for excision in the temporal domain and is useful for impulsive and periodic RFI. Lost data in this method is counted as a loss of observing time.

RFI pre-detection and suppression methods in the data reception phase require real-time operation. This is not expected to be a big problem for single telescopes, but will be a challenge especially for large telescope arrays. A systematic study of the RFI environment including the implementation of a real-time RFI mitigation system is described in Baan et al.\upcite{11} using a trial experimental Field-Programmable Gate Array (FPGA) setup. The FPGA performs an initial frequency domain filtering followed by the thresholding of received data using cumulative summing. The application is illustrated in Fig. 5. Similar experimental studies will help in characterization and excision of RFI at early stages such that implementation can be built into the telescope array setup, possible with upcoming observational facilities, such as Square Kilometre Array (SKA).
%ͼ5
\begin{figure}[tbph]
\label{pic5}
\centering
{\includegraphics[angle=0,width=9cm]{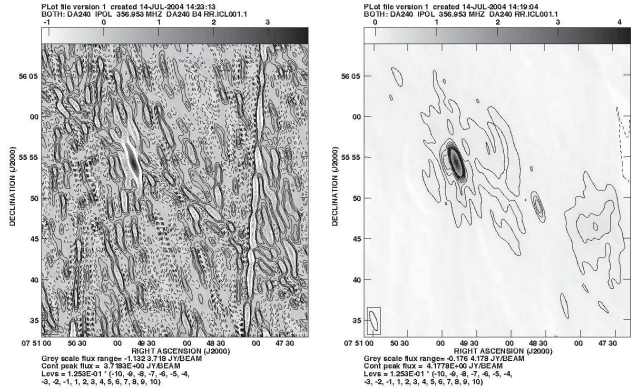}}
%\vspace{-2mm}
\caption{Continuum emission from the giant double radio galaxy DA240 at 357 MHz before (left) and after (right) the removal of RFI. The contrast between these two images is obvious. After RFI excision, the image quality has been significantly improved. Image source: Baan et al.\upcite{11}.}
    \end{figure}

\subsection{Pre-correlation strategies}

These strategies are applied to either the received raw data or that output from the pre-detection filtering before correlation and further processing, functioning as a means to identify and reduce RFI contamination in near real time. They can be implemented through spatial excision which refers to cancellation of the RFI signals at the antenna array level, as an adaptive noise cancellation (ANC) scheme in the temporal domain, and methods relevant to the temporal blanking of data reception. We discuss these techniques below.
\begin{itemize}
\item Spatial excision at the antenna and instrumentation level are RFI mitigation techniques during the initial imaging/beamforming stage are applicable to sources of RFI whose locations and directions are known. In single-dish telescopes, astronomical and RFI signals are coherently added, while in interferometric arrays RFI partly enters the receiving system incoherently, enabling an in-built strategy. Extended baselines thus serve as a spatial filter of RFI. Nevertheless, calibration of each station in an array is still affected by its local RFI.

Spatial excision can be addressed in multi-antenna systems by adjusting the beam pattern such that the nulls correspond to positions occupied by strong RFI sources (e.g. Ellingson \& Hampson\upcite{37}; Hansen et al.\upcite{50}; Landon et al.\upcite{65}); similar to subspace projections where an eigenvalue decomposition (singular value decomposition) of the covariance matrix composed of signals received at differing times by each component of an interferometric array is used to eliminate RFI (e.g. Leshem et al.\upcite{67}; Briggs \& Kocz\upcite{26}; Boonstra \& van der Tol\upcite{21}; Pen et al.\upcite{95}; Kocz et al.\upcite{61}; Paciga et al.\upcite{91}); and the identification and excision of RFI due to telecommunication or other transmissions where it is known that some statistical property (such as their mean, median, correlation, etc.) of the signal is periodic (cyclostationary, non-stationary process, e.g. Weber et al.\upcite{116}; Hellbourg et al.\upcite{54} using a phased array).

\item Waveform filtering or ANC is a means of filtering temporal RFI through thresholding using convolution kernels in the frequency domain and returning back to the temporal domain before the data is fed to the correlator. Commonly used methods employ the Wiener filter (e.g. Barnbaum \& Bradley\upcite{15}) for known sources of RFI and the assumption that the RFI source is stationary. As the procedure of filtering essentially acts on a time ordered set of data, they are applicable to continuum observations and pulsar search and monitoring studies (e.g. Kesteven et al.\upcite{58}).

\item Temporal blanking technique is the halting of the correlator input during strong RFI reception durations and the associated study of the variance properties of the received batch samples (e.g. Weber et al.\upcite{115}; Fridman\upcite{41}). Clipping based techniques, which includes the thresholding of a signal if the cumulative sum of the preceding signals exceed a pre-defined threshold (e.g. Baan et al.\upcite{12}; Fridman\upcite{42}) complement the blanking technique. The identification of RFI is based on the systematic deviation from the expected $\chi^2$ distribution of random Gaussian noise signals and higher order statistics in the accumulating dynamic power spectrum (e.g. Fridman\upcite{40}), and deviation from expected statistics of random Gaussian noise in the time domain\upcite{13}. The measurement of spectral kurtosis from the accumulated power spectra and their square can be used to quantify deviation of a received signal from expectations due to Gaussian random noise\upcite{81,46,80}.

\end{itemize}

\subsection{Post-correlation strategies}

These strategies are applied to post-correlated data and involve the handling of either time slices of data at the observation frequencies, two dimensional time-frequency data or integrated data cubes such that the operation is performed before or during the imaging stage. A general review of some of these strategies with a focus on accurate flagging and thresholding by accounting for the data statistics is presented in Offringa et al.\upcite{84} Traditional post-correlation processing consists of flagging and excision, which is time consuming and often done manually, and is thus is not suitable for the processing of future large-scale observation data such as with the SKA. In addition, as these traditional methods handle integrated and correlated data, this will lead to a significant data loss and affect the overall quality of the data, 
and hence the identification of useful astronomical signals.

Post-correlation RFI mitigation can be implemented through some methods also applicable as pre-correlation strategies, such as smoothing and surface fitting, fringe stopping based identification, and thresholding. We discuss these techniques below.
\begin{itemize}
\item Digital filtering of signals outside the given target frequency range of the receiver (e.g. Ellingson et al.\upcite{36}; Haoran Li et al.\upcite{129}) employs window functions in the Fourier frequency domain, ensuring the preservation of the signal arrival information. Spatial nulling and the adaptive cancellation of RFI\upcite{130} using a reference channel or antenna pointed towards the known strong RFI source can be used to subtract the RFI signal from the primary phased array-based astronomical signal (e.g. Barnbaum \& Bradley\upcite{15}; Briggs et al.\upcite{25}; Ellingson \& Hampson\upcite{37}), and can be simulated to be applicable as both pre- and post-correlation strategies\upcite{76}.

\item Surface fitting involves the fitting of the time-frequency data plane using either pixel slices, smooth functions (e.g. as discussed in Offringa et al.\upcite{84}) or in the Fourier domain\upcite{118} for image difference based RFI flagging and the possibility of image reconstruction (e.g. Winkel \& Kerp\upcite{117}).

\item The effects of fringe stopping picked up by constant, stationary sources of RFI (with respect to the beam phase center; varying with different baselines\upcite{9}) can be accounted for and subtracted. An algorithm involving this method along with thresholding in the Fourier domain is integrated into the Astronomical Image Processing System (AIPS) pipeline\upcite{62}.

\item Cumulative summing based thresholding for a set of sequential samples, similar to the earlier described pre-correlation strategy of temporal blanking, can be employed here to take advantage of an accurate time stamping and hence RFI mitigation offline.

\item Combinatorial thresholding of sequentially or morphologically related samples can be employed in RFI mitigation. The methods effectively measure the level of connectedness to assign weights to RFI contaminated data portions and based on a pre-defined threshold, flag those portions (e.g. Offringa et al.\upcite{84,86}). Automatic flagging and thresholding methods are often applied to pipeline processing systems of interferometric arrays. Some examples include Pieflag (used for the Australia Telescope Compact Array, ATCA; Middelberg\upcite{73}), AOflagger (used at the Low-Frequency Array, LOFAR and at the MWA; Offringa et al.\upcite{83,86,89}), that developed for the Allen telescope array (e.g. Keating et al.\upcite{57}), FLAGCAL (used at the Giant Meterwave Radio Telescope, GMRT; Prasad \& Chengalur\upcite{98}), and Scripted E-merlin Rfi-mitigation PipelinE for iNTerferometry (SERPENT) which is an RFI thresholding method used to calibrate the electronic Multi-Element Radio Linked Interferometer Network, e-MERLIN\upcite{94}.

One must be careful using automated thresholding techniques, as these flaggers could potentially identify real signals less than 2 minutes as RFI contaminants. An example is the case involving the TDE source Swift J1644+57 with LOFAR (e.g. Cendes et al.\upcite{29}), where the source was not detected, and possibly identified as RFI owing to its weak transient signal.
\end{itemize}

Thresholding techniques must thus be supplemented by the use of the expected statistical distribution of data and random Gaussian noise to infer deviations due to RFI contamination, similar to the methods described earlier. Advantages of post-correlation analysis include the handling larger volumes of data (including initial likelihood based flagging and processing) and the possibility of using information on the accumulated data statistical properties in the training of automated RFI flagging and excision algorithms with minimum human intervention. Such attempts have been made recently with the training of simulated post-correlated data using a neural network (an approach towards deep learning) and the application of thresholding algorithms\upcite{3} to test data with compelling results in terms of RFI identification accuracy, though with a requirement for faster and less computationally intensive approaches, thus still not applicable to large scales.

\section{Science cases of RFI suppression and mitigation strategies in radio observatories}

Astronomical signals of interest include visibility data from interferometers and flux density measurements from single antenna, both of which are distributed in the time-frequency domain in a manner depending on the source and the observational capacity of the facility. Visibility data is composed through interference synthesis of a multiple telescope interferometric array. Compared to single antenna, imaging sensitivity and resolution of visibility data are greatly improved, and can be used to study the formation of stars in the Milky Way and the late stages of stellar evolution, accurately measure the location and distance of celestial sources of maser emission, the physical properties of pulsars and precision ranging, active galactic nuclei (AGN) activity, starburst activity during galaxy mergers and interactions, galaxy clusters, surveys of HI spanning the Galaxy, and on cosmic scales which aid studies of the epoch of reionization, dark matters through baryon acoustic oscillations, establishment of angular accuracy of the celestial reference frame, monitoring the Earth plate movement and so on. The pulsar timing data of single antenna is used to re-construct unique pulse profile shapes of known pulsars, discover new pulsars and other repeating transients (possibly rotating radio transients and fast radio bursts) from surveys, and employ the accurate timing information in the setup of pulsar timing arrays which are aimed at the detection of the gravitational wave background from coalescing supermassive black holes. A recent review of the development of extragalactic radio astronomy is presented in Padovani\upcite{92}. Early scientific goals of the upcoming Square Kilometre Array (SKA) are presented in Carilli \& Rawlings\upcite{28}. The latest scientific white paper of SKA contains updates on key scientific objectives and research areas. The SKA scientific conferences held in Sweden in 2015, and in India in 2016 provide platforms for discussing the scientific program spanning the first phase of SKA (SKA1). With recent widely participated meetings, Chinese researchers are currently in the process of compiling a Chinese SKA science white paper and science program.

Of particular note is that all the above-mentioned advances in astronomical research are inseparable from progress of technology, including the development of new hardware devices (new antenna and broadband feed, low temperature amplifiers of receivers, high-speed electronic recording terminals, software and hardware correlators), and the development of automatic pipeline processing software system adapted to the large-scale observation data, and analysis of large data sets (data processing, classification, storage and retrieval, user interactivity, etc.). The quality of the observed data strongly influences scientific results and the authenticity of the inferences. Radio frequency interference suppression and mitigation technology is also progressing and developing under the guidance of scientific needs. Below, we present the major scientific directions and the RFI protection, suppression and mitigation methods adopted by some representative observational facilities at home and abroad (focused mainly on low frequency interferometric arrays) as the experience gained from their implementations will help identify areas for improvement, especially in the development of tailored strategies for RFI mitigation.

\subsection{21 cm Array (21CMA)}

The 21CMA is located on the plateau of Ulastai, deep in Tianshan, Xinjiang of west China, is an indigenously constructed facility. It is the first dedicated radio interference array to aid in the detection of ``the first ray of the universe''. The array consists of 81 pods with a total of 10287 log periodic antennas operating in the 70 -- 200 MHz frequency range and will be able to probe the neutral hydrogen 21 cm emission in the z = 6 - 27 redshift range, to study the formation and evolution of the first generation of luminescent celestial sources during the epoch of reionization (cosmic microwave background, CMB)\upcite{120}.

The identification, characterization and excision of RFI at the 21CMA site are detailed in the study of Huang et al.\upcite{55} The main sources of RFI are identified as due to civil aviation communications at 119 MHz and 130 MHz, ORBCOMM communication satellites based signals in the 137 - 138 MHz frequencies (see Table \ref{Table:RFISource}), walkie-talkie based communication signals from passing trains at 151 MHz, the FM radio transmission band, and internally generated RFI from the instrumentation and computing resources. RFI mitigation is addressed in several stages so as to minimize its influence. Interferometric correlations and long integration times are used to identify RFI through its deviation from the Gaussian probability distribution function. Strong RFI signals can be easily identified in frequency spectra. The suppression of time-varying RFI is then performed through the handling of weighted visibilities to minimize its impact.

\subsection{Five-hundred-meter Aperture Spherical Telescope (FAST)}

The FAST is currently the largest single dish telescope, constructed indigenously and spans a diameter of 500 m. It is situated in Guizhou province of China, operational in the 70 - 3000 MHz low-mid frequency range. Science capabilities include deep HI surveys including in the Milky Way and in extragalactic sources, pulsar general investigations and searches, molecular line and maser studies (e.g., hydroxyl) amongst others (e.g. Nan et al.\upcite{79}; Li et al.\upcite{68}; Li \& Pan\upcite{69}; Pan et al.\upcite{125}).

FAST is susceptible to RFI owing to its high sensitivity. Measures to prevent and eliminate interference affecting high sensitivity are a key part of FAST interference research. The natural setting for FAST in a karst depression in a remote site 
enables minimum RFI contamination (e.g. Nan et al.\upcite{79}). Measures taken for RFI mitigation at the observatory level include the Guizhou government regulated setup of a radio-quiet zone with a 30 km radius around the telescope site, establishing regulatory facilities to maximize the protection of the current and future radio environments (Figure 4), and an EMC study for the identification and regulation of the use of instrumentation and computational resources, causing minimum internal RFI\upcite{122}.  
Additional interference mitigation related to operations of FAST are currently underway.  
Strategies of interest could involve pre-correlation techniques including thresholding and median filtering to handle timing and spectral data such as the study of Bhat et al.\upcite{20} which was applied to test data from Arecibo and the Green Bank telescope; or pipelines for the identification of discriminating statistical properties of Gaussian noise from that due to RFI (e.g. Andrecut et al.\upcite{6}). Since FAST has high sensitivity and can easily identify RFI signals, methods that proposed for temporal blanking can also be used.

\subsection{Tianma Telescope}

The Tianma telescope, which also called Shanghai 65 meters radio telescope, is located in Sheshan, Songjiang District, Shanghai. Currently it is the second largest radio telescope in Asia in addition to FAST, and is also an important component of the international VLBI network. It is operated by the Shanghai Observatory, Chinese Academy of Sciences. The telescope uses a Cassegrain Reflector Antenna with a wide range of operating bands, with receiving devices distributed in 8 bands in L, S, C, X, Ku, K, Ka, Q\upcite{131}. It is the first radio telescope in China operational in the 7 mm wavelength. Scientific research includes molecular spectrum observations (including the spectra of massive stellar formation region, K-band galactic plane spectra, spectra of high red shift carbon monoxide and other important molecular; Wu\upcite{132}), and studies of pulsars and other radio sources (including X-ray binaries, blazars, etc.)\upcite{133}.

The RFI contamination at the Tianma telescope is mainly in the C-band. Mitigation strategies include the establishment of a radio quiet area of 3 km, monitoring RFI in the frequency range 1 - 12 GHz, estimating the devices with active surface consisting of 1200 triggers, removing strong interference in the S and L bands using superconducting filter, and for RFI pollution within the channel, directly removing them in the terminal. These strong interference excised data are then used for scientific research.

\subsection{e-MERLIN}

The e-MERLIN, situated in the United Kingdom which is operational in the 151 MHz - 24 GHz (Garrington et al.\upcite{45}) is a technologically upgraded version of the MERLIN array in terms of receivers, analog and digital electronics and optical fibers (instead of transmitting data by microwave; e.g. Norris et al.\upcite{82}). MERLIN and e-MERLIN have been employed in many science studies including to trace stellar mass loss during their formation and late stage evolution (e.g. Richards et al.\upcite{101}; Ainsworth et al.\upcite{2}; Morford et al.\upcite{77}), observations of the pc- to kpc-scale jet features of AGN (e.g. An et al.\upcite{4,5}; Gabanyi et al.\upcite{44}; Araudo et al.\upcite{7}; Straal et al.\upcite{108}), radio bright outbursts in other galaxies (e.g. Argo et al.\upcite{8}), and to distinguish AGN from starburst activity by inferring atomic and molecular line based diagnostics including the expected optical depth and inflow/outflow behavior\upcite{31}, targeted and surveys of maser emission (OH, water, methanol, etc.) from gas clouds and star forming regions and evolved stellar envelopes (e.g. Green et al.\upcite{48}; Etoka et al.\upcite{38}; Richards et al.\upcite{100}; Wolak et al.\upcite{119}). An additional interesting source of intensive study includes searches for progenitors and associated sources for fast radio bursts (FRBs), including the recently observed FRB 150418 (e.g. Bassa et al.\upcite{16}; Giroletti et al.\upcite{47}).

The typical level of RFI contamination measured is $(5-25)\%$\upcite{94} depending on the particular receiver. The existing RFI excision is based on basic interferometry, and since the upcoming facilities (technological and scientific goals) will be more complex, the study of Peck \& Fenech\upcite{94} extended the post-correlation strategy used at LOFAR for use by e-MERLIN (SERPENT), including the use of combinatorial thresholding and cumulative summing techniques. The RFI mitigation pipeline used at the e-MERLIN is presented in Fig. 6.
%ͼ6
\begin{figure}[tbph]
\label{pic6}
\centering
{\includegraphics[angle=0,width=9cm]{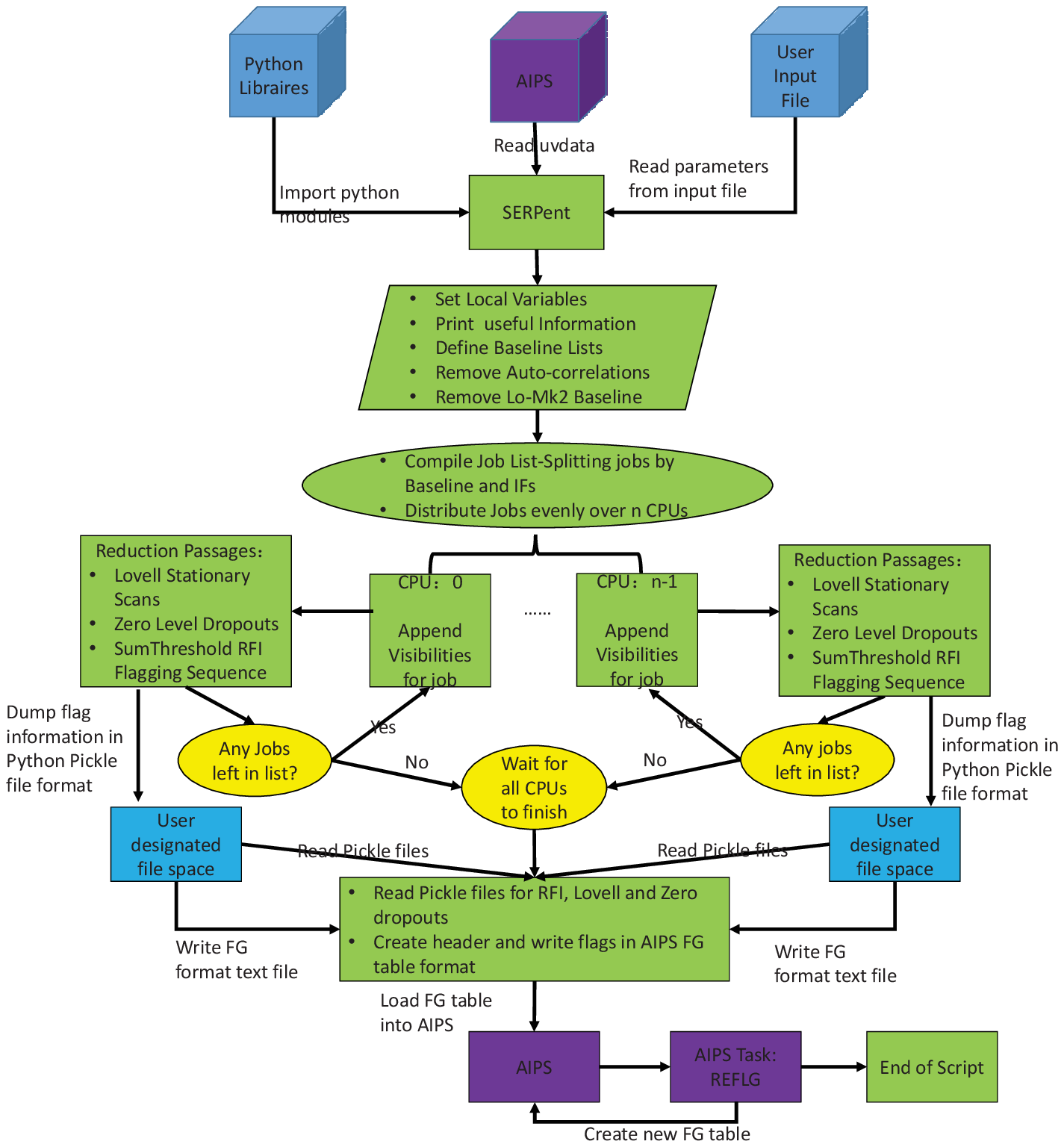}}
%\vspace{-2mm}
\caption{Suggested parallelized RFI mitigation pipeline for e-MERLIN by using combinatorial thresholding and the median absolute deviation based filtering. (Reproduced from Peck \& Fenech\upcite{94})}
    \end{figure}

\subsection{GMRT}

The GMRT, situated near Pune, India is operational in the 50 - 1420 MHz low frequency range. Scientific studies cover a vast range of areas including pulsar search (flux density measurements, spectral indices) and timing (e.g. Gupta et al.\upcite{49}; Dembska et al.\upcite{34}), galactic and extragalactic spectroscopic and continuum observations through targeted observations and all sky surveys (e.g. Roy et al.\upcite{104}), study of molecular clouds and star forming regions (e.g. Ainsworth et al.\upcite{1}), studies of radio galaxies and AGN involving jet morphology and identification or resolving of dual core sources (e.g. Kharb et al.\upcite{59}; Singh et al.\upcite{107}), galaxy clusters, their kinematics and activity based signatures (e.g. Lindner et al.\upcite{70}), and searches for the epoch of reionization (e.g. Pen et al.\upcite{95}; Paciga et al.\upcite{91}), amongst others.

The radio environment at GMRT is mainly contaminated by FM transmission and unshielded power cables and TV signals, many of which have been identified and subjected to corrective action. This includes the effort to successfully introduce government regulated measures to switch the operation of the GSM network from the existing 850/900 MHz to 1800 MHz\upcite{99}. RFI mitigation has been carried out as a pre-correlation strategy using the singular value decomposition method and by the tracking and interferometric excision such as in the epoch of reionization studies\upcite{91}; and the use of fringe stopping correction to excise RFI which does not produce fringe patterns\upcite{9} as discussed briefly in post-correlation strategies; the developed signal processing pipeline\upcite{103}, also presented in Fig. 7, flags and excises RFI by blanking of received voltage streams at the pre-correlation stage above a pre-defined threshold based on the median absolute deviation estimator.
%ͼ7
\begin{figure}[tbph]
\label{pic7}
\centering
{\includegraphics[angle=0,width=9cm]{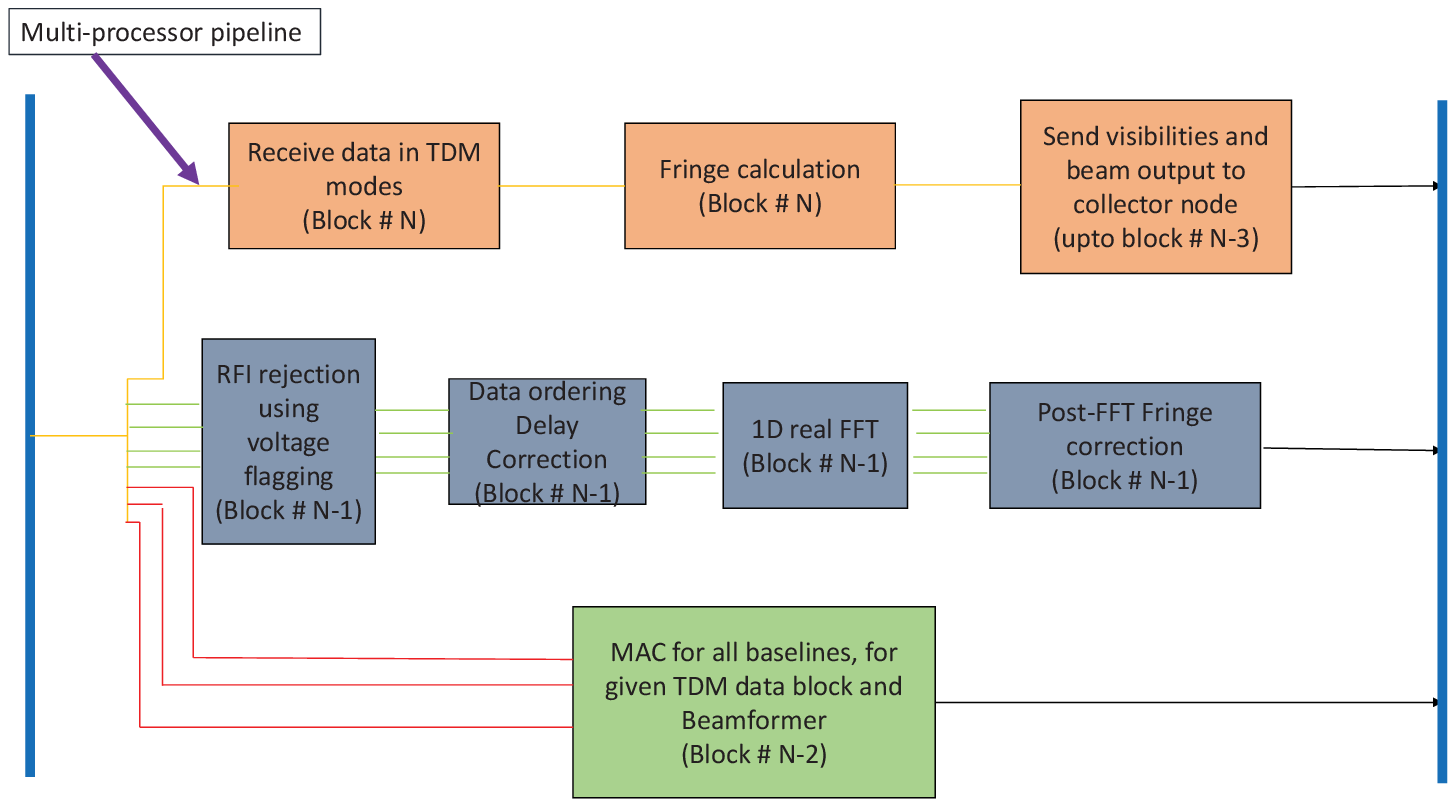}}
%\vspace{-2mm}
\caption{Data processing pipeline for a single node and the stage illustrating the RFI excision for the pulsar timing data processing at the GMRT. (Reproduced from Roy et al.\upcite{103})}
    \end{figure}

\subsection{LOFAR}

The LOFAR, situated mainly in the Netherlands with additional stations distributed over other European countries (Germany, France, Sweden and the United Kingdom) is operational in the 30 - 80 MHz and 110 - 240 MHz frequency ranges. The vacancy of the intermediate frequencies is to circumvent the typical FM radio channels. Areas of study include the search for the epoch of reionization\upcite{56}, transient searches and monitoring\upcite{63,97}, cosmic magnetism\upcite{18}, galactic structure formation through the collapse and evolution of gas clouds, nearby and distant galaxies\upcite{32,78}, galaxy cluster\upcite{111}, cosmic rays\upcite{106} and solar physics, amongst others in both targeted and survey modes (e.g. van Haarlem et al.\upcite{110}, and references therein).

Studies of the RFI environment at the LOFAR sites indicate that persistent sources include FM radio and TV transmissions\upcite{19}, mainly narrow band contamination with an occupancy of (1.8 - 3.2)\%\upcite{88}. The empirical estimation of both weak and strong RFI contamination account for major sources, especially for the sensitive epoch of reionization studies \upcite{87}. The mitigation of RFI has been addressed mainly as a post-correlation strategy by the use of combinatorial thresholding\upcite{84} and the software package AOFlagger\upcite{86} which can handle data at different resolutions in a scale-free manner. The post-correlator data is hence subjected to automated flagging and excision of RFI. The flagging pipeline used at LOFAR is presented in Figs. 8 and 9.

%ͼ8
\begin{figure}[tbph]
\label{pic8}
\centering
{\includegraphics[angle=0,width=9cm]{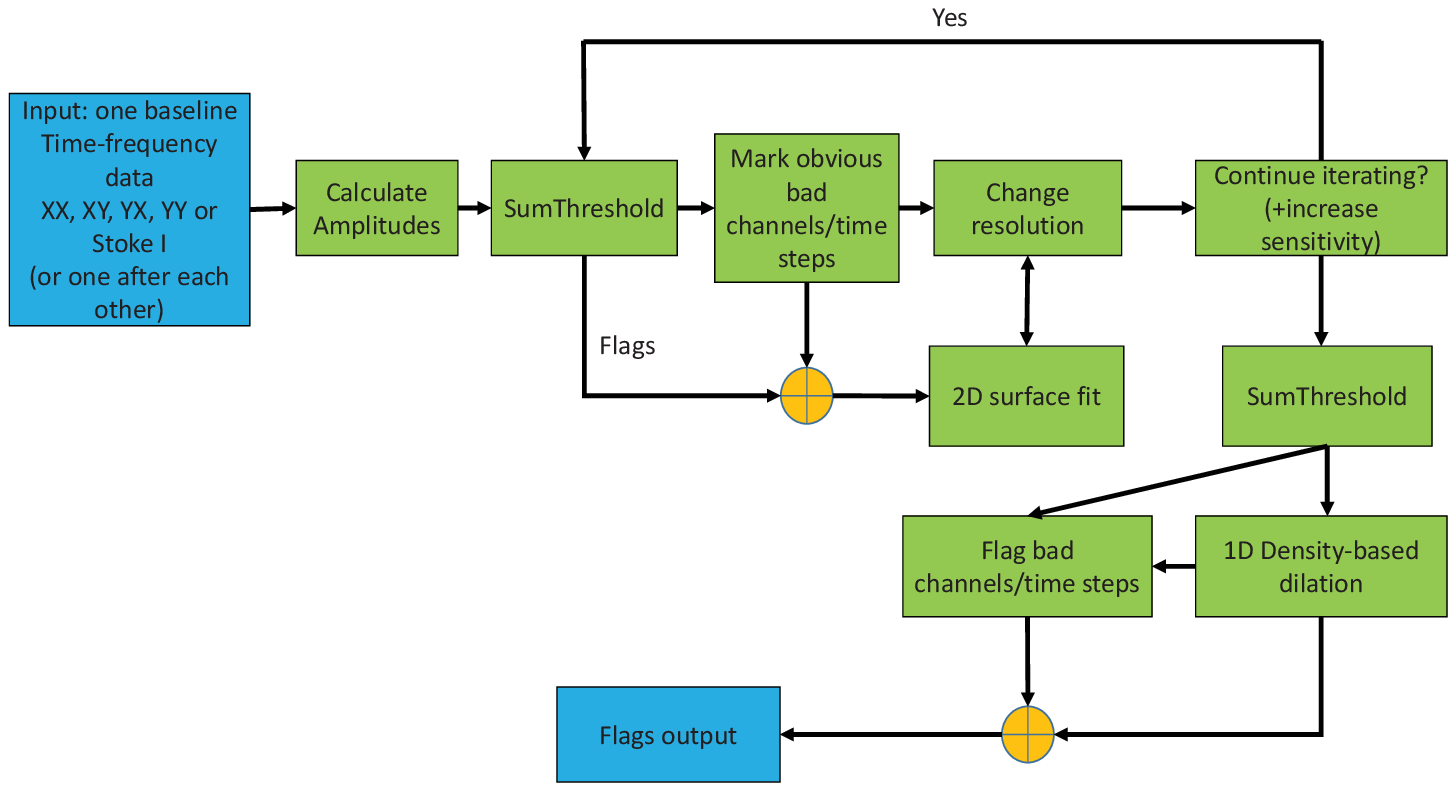}}
%\vspace{-2mm}
\caption{RFI flagging pipeline tested and adopted at LOFAR using the combinatorial thresholding method. (Reproduced from Offringa et al.\upcite{85})}
    \end{figure}

%ͼ9
\begin{figure}[tbph]
\label{pic9}
\centering
{\includegraphics[angle=0,width=9cm]{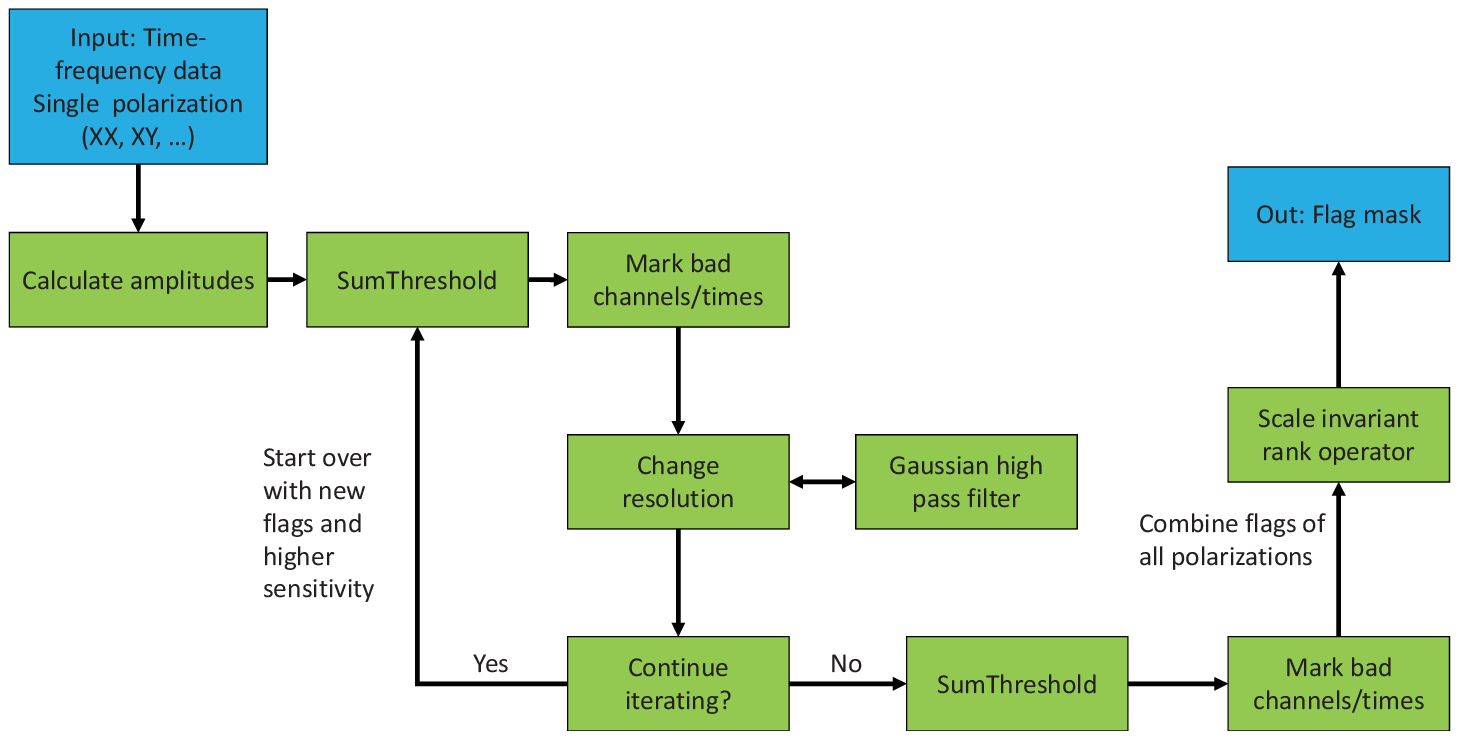}}
%\vspace{-2mm}
\caption{Updated RFI flagging pipeline at LOFAR using the combinatorial thresholding method and the scale invariant rank operator. (Reproduced from Offringa\upcite{83})}
    \end{figure}

\subsection{WSRT and APERture Tile In Focus (APERTIF)}

The Westerbork Synthesis Radio Telescope (WSRT), situated in the Netherlands and operated by the Netherlands Institute for Radio Astronomy (ASTRON) is operational in the 120 MHz - 8.3 GHz frequency range and contains 14 telescopes arranged along the east-west direction. Areas of study include HI surveys based imaging of galaxies (e.g. Braun et al.\upcite{24}; Oosterloo et al.\upcite{90}; Heald et al.\upcite{52}) and contribution to other surveys such as the Arecibo Legacy Fast Arecibo L-band Feed Array (ALFALFA)\upcite{66}, targeted spectroscopic observations of OH absorption line systems (e.g. Vermeulen et al.\upcite{113}), polarization and Faraday rotation measurements to constrain magnetic field strength and topology in galaxies (e.g. Heald et al.\upcite{51}; Braun et al.\upcite{23}), targeted and surveys of galactic and extragalactic maser emission (water, formaldehyde, methanol, etc.), atomic and molecular lines from star forming regions, dust and gas clouds (e.g. Dickel et al.\upcite{35}; Klockner \& Baan\upcite{60}; Yim \& van der Hulst\upcite{121}).

The APERTIF is a focal plane array prototype operating in the 1000 - 1700 MHz frequency range\upcite{112}, acting as an extension of the WSRT with regard to a larger field of view and bandwidth, thus aiding better focus on the HI and OH surveys, and in pulsar searches\upcite{102}.

The pre-correlation strategies including blanking, cumulative summing and inference and subsequent excision by inference of deviations from expected statistical properties of the received data and noise are identified as effective in performing RFI mitigation in real-time (e.g. Fridman\upcite{40}; Baan et al.\upcite{12,11}).

\subsection{SKA precursors/pathfinders}

The two SKA sites located in Australia and South Africa include a low-frequency array and mid-frequency arrays respectively. These include four SKA precursors: the MWA and Australian SKA Pathfinder (ASKAP) in Western Australia, and MeerKAT and Hydrogen Epoch of Reionization Array (HERA) in South Africa. For the MWA and ASKAP, preventive strategies are in place owing to the suitable selection of host sites and their preparation. A detailed review of continuum surveys with current and proposed SKA pathfinders is presented in Norris et al.\upcite{82}.

For the ASKAP, operational in the 700 - 1800 MHz frequency range, pre-correlation strategy involving spatial filtering using a reference antenna approach is tested and was suggested for use\upcite{53}. Recently, the singular value decomposition method was suggested based on its application to the initial Boolardy Engineering Test Array (BETA;  McConnell et al.\upcite{72}) consisting of six antenna to handle RFI from satellite communications\upcite{14}\footnote{http://www.atnf.csiro.au/projects/askap/ACES-memos}.

The MWA is a SKA precursor operational in the 80 - 300 MHz frequency range. It is situated in Murchinson outback of Australia, a site preserved through governmental regulations to be a radio-quiet zone and is focused mainly on the search for the epoch of reionization, all sky surveys (Galactic and extragalactic), transient studies and solar physics\upcite{71,109,22}. The blanking technique is found effective in the reduction of RFI, especially in the FM radio bands\upcite{75}. The AOFlagger based RFI flagging and thresholding algorithm has been adopted for application to the MWA data (e.g. Offringa et al.\upcite{89}; Beardsley\upcite{17}), and indicates less than 3 \% contamination in data (mainly due to FM radio and digital TV signals) owing to the remote location of the observatory. The former study also advocates the development of pre-correlation hardware interfaces (at the receiver), and a high time-frequency resolution for near real-time RFI excision. It is thus expected that a combination of both pre- and post-correlation techniques will be effective in the flagging and excision of RFI.

The MeerKAT is a SKA precursor currently in the development and testing phase which will be integrated into the SKA-1 Mid array upon expansion. The core of the instrumental array will be situated in the Karoo desert of South Africa, a site identified as a radio quiet zone after extensive site surveys. The KAT-7 is a compact antenna array operational in the 1200 - 1950 MHz, acting as a testbed for science and technological possibilities with the MeerKAT\upcite{27}. Science studies addressed using KAT-7 including HI observations of low brightness nearby galaxies and clusters, transient sources including novae, X-ray binaries, AGN, and pulsars, integration into VLBI observations, and OH maser studies (Foley et al.\upcite{39}, and references therein). RFI excision in HI studies of the gas rich nearby galaxy NGC 3109 was carried out using a low bandpass filter, and was identified as originating due to internal instrumental sources including the antenna\upcite{27}. It is also suggested that the shielding of the digitizer associated with the antenna can be used in further reducing instrumental RFI \upcite{39}.

The HERA is the latest defined SKA precursor and is focused on deep HI observations in the redshift range z = (6 - 12) in order to study the epoch of reionization and the cosmic dawn period at z $ \approx $ 30\upcite{33}, covering similar objectives as the 21CMA and other experiments such as the Precision Array for Probing the Epoch of Reionization (PAPER; Parsons et al.\upcite{93}) in South Africa. The radio interference environment for these facilities is different necessitating a tailored approach. Though, as the science goals are similar, the mitigation of RFI in stages such as that used in the 21CMA and MWA is definitely useful.

\section{Summary}

The purpose of this paper is to introduce the commonly used RFI suppression and mitigation strategies, to determine their current usage and status, and to introduce the latest advances in related technologies. Understanding the motivations and usage scenarios of these strategies is very useful for their implementation, improvement, and the development of new methods. The adoption of appropriate RFI mitigation techniques for a given observational facility (single dish or interferometric array) must be made after an assessment of the particular RFI environment, the observation mode (timing; imaging: continuum, spectral line), compatible hardwares and softwares, and available computing resources, amongst others. Challenges expected to be faced include the requirement to handle large data volumes resulting from higher bandwidth, higher time and frequency resolution of existing and upcoming observational facilities, and the ensuing non-linearity of the subsequent mitigation.

Rapid advances in computing (processing, storage, retrieval and user interactivity) including the use of off-line clusters and supercomputers and online cloud computing allows for a consistent implementation of RFI mitigation at both the pre- and post-correlation stages. Any new technological advance can thus be absorbed within the existing work flow without additional time and costs involved in the modification, testing and upgrading of instrumentation and software pipelines involved in the pre-correlation setup of existing observational facilities.

A quantitative assessment of the level of mitigation required may be made specific to the observatory in order to deploy a suitable form of computational resource and combined strategy. In addition, a coordinated means for spectrum management (definitions, implementation and maintenance) which is objective and addresses concerns for all involved parties is required. This can only be achieved by properly accounting for existing devices and self- regulation so as to improve and define better capabilities for astronomy observations by minimizing the influence from increasingly wide-spread unlicensed (commercial, communication) contaminants. Here, an overview of the main strategies was presented, specifically in the context of astronomical research where the dominant terrestrial transmissions are detrimental to the reception of astronomical signals.  The implementations by some observing facilities (mainly low frequency) serves to illustrate the applicability of these strategies and as possibilities for future use.

It must be stressed that preventive strategies which pro-actively tackle incoming RFI signals are much better than mitigation as they reduce data contamination, and result in lesser data loss and accumulating costs involved in subsequent mitigation due to the deployment of additional computational resources. Pre-correlation blanking and filtering performed in real time can result in a considerable suppression of the RFI signals. The implementation of these techniques relies on fast data processing, which was once a limiting factor to its adoption. However, this is now much easier and technically feasible owing to the rapid development of modern digital signal processing devices such as FPGAs. Real-time post-correlation is performed during the correlation process but before data averaging. It utilizes the statistical properties of the data and can be effective especially for interferometers, but it may also require special hardware to handle large incoming data volumes. The computational runtime of any real-time RFI mitigation technique is mainly dominated by the reading and writing of the data in addition to the processing time. Thus, the input-output (I/O) disk performance of the data processor (e.g., a cluster) greatly influences the effectiveness of the implementation. Investigations of parallel I/O approaches made by the SKA data management research team (e.g., Wang et al.\upcite{114}) can offer beneficial ideas for improvements in the performance of large scale RFI mitigation strategies. Though off-line post-correlation approaches may be less effective than those in real-time, they are more flexible and hence practically attractive. They do not require changes to the existing front- and back- end devices, and are compatible with the widely employed software correlators. Automated off-line flagging is also necessary for dealing with weak RFI. The increased interference-to-noise ratio after data integration is useful for detection of weak RFI which may be neglected from real-time flagging.

There is a requirement for the restoration of useful data after the identification and removal of RFI, motivating automated methods of classifying a signal as RFI based on statistical considerations to reduce false positive detections. This is also motivated by the increasingly large data volumes from existing and upcoming observational facilities such as FAST, SKA precursors (MWA, ASKAP, MeerKAT and HERA) and pathfinders (e.g. 21CMA, GMRT, LOFAR). It is then necessary to introduce, test and deploy more sophisticated algorithms than simple flagging and thresholding. A practical manner of RFI mitigation could be the use of a combination of various techniques in the causal stages, enabling the considerable suppression of RFI. Certain strategies also enable the partial restoration of data, making it possible to fully utilize historical archival data for training the algorithm, testing and subsequent live deployment.

The RFI monitoring experience from the MWA and MeerKAT implies that SKA (which will be constructed at the same sites as these precursors) will be faced with an unavoidable RFI contamination mainly from the aircraft and satellite transmissions, even after the adoption of regulatory preventive measures. Due to the high sensitivity of the SKA and the demand of extremely high dynamic range for EoR study, low-level RFI will present a big challenge with the requirement to handle it with the highest priority. Figure 10 demonstrates a flow chart of the suggested data inspection and pre-reduction including RFI mitigation component. The advantage of high time and frequency resolution of the SKA enables accurate identification and excision of RFI contamination, thus decreasing the data loss; however to deal with a large data volume of the SKA scale, a huge computational power is required to perform the RFI excision. In addition to computing challenging, the available buffer of the SKA science data processor sets another limitation to RFI detection since in the process of RFI identification, intermediate data are to be kept in memory, which requires same-size memory as input snapshot data set\upcite{89}. Current RFI mitigation software packages are to be parallelized on multiple nodes in high performance computing (HPC) platform to relieve the pressure on computing and storage, though, the current low execution efficiency of astronomical softwares and algorithms on HPCs (10\%-25\%) are still a concern.

%ͼ10
\begin{figure}[tbph]
\label{pic10}
\centering
{\includegraphics[angle=0,width=9cm]{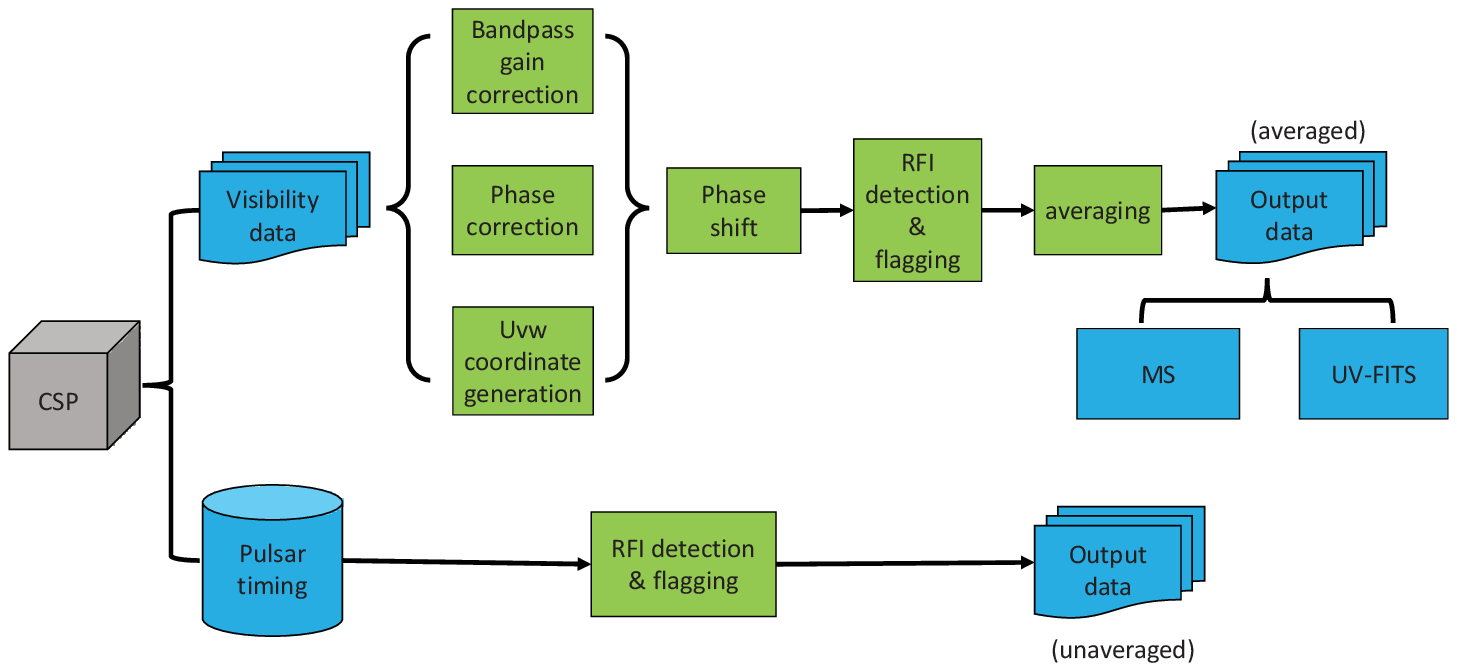}}
%\vspace{-2mm}
\caption{Sketch map of the SKA data inspection and pre-processing including RFI mitigation step.}
    \end{figure}

\acknowledgements{This work has been supported by the National Key Research and Development Program of China (2016YFE0100300), Strategic Priority Research Program of the Chinese Academy of Sciences (CAS; Grant No. XDB23000000 ).
Tao An thanks the grant support by the Youth Innovation Promotion Association of CAS. P. Mohan acknowledges support through the CAS President's International Fellowship for Postdoctoral Researchers program (PIFI: 2016PM024) and the National Natural Science Foundation of China (grant no. 11650110438). The authors thank Willem Baan for their useful discussion of the draft of the paper, thank Rendong Nan and Haiyan Zhang for providing information on RFI mitigation strategies of FAST, and Bin Li for providing the RFI reduction information of Tianma telescope. Finally, we are grateful to Xiangping Wu on the constructive comments on the overall RFI project.}

\end{document}